\documentclass[aps,prl,reprint,superscriptaddress]{revtex4-2}

\usepackage{graphicx}    % For figures
\usepackage{amsmath}     % For math (includes amssymb, amsfonts)
\usepackage{bm}          % For bold math
\usepackage{siunitx}     % For SI units
\usepackage{xcolor}      % For color comments/annotations

\definecolor{seagreen}{rgb}{0.18, 0.55, 0.34}
\usepackage{hyperref}
\hypersetup{
    colorlinks=true,
    linkcolor=seagreen,      % internal links
    urlcolor=seagreen,       % URLs
    filecolor=seagreen,
    citecolor=seagreen,      % 
    linktoc=all
}
\usepackage[normalem]{ulem}
\usepackage[margin=1in]{geometry}
\usepackage{booktabs}
\usepackage{subcaption}

\begin{document}

\title{The diffusion coefficient in the Large Magellanic Cloud}

\author{J. Reynoso-Cordova}%\email{...}
\affiliation{Department of Physics, University of Alberta,
CCIS 4-181, Edmonton, Alberta T6G 2E1, Canada}
\affiliation{Istituto Nazionale di Fisica Nucleare, Sezione di Napoli, 
Complesso Universitario di Monte Sant'Angelo, 
Via Cintia, 80126 Napoli, Italy}

\author{D. Gaggero}%\email{...}
\affiliation{INFN Sezione di Pisa, Polo Fibonacci, Largo B. Pontecorvo 3, 56127 Pisa, Italy}

\author{M. Regis}%\email{marco.regis@unito.it}
\affiliation{Dipartimento di Fisica. University of Torino, \textit{I-10125}, Torino, Italy}\affiliation{Istituto Nazionale di Fisica Nucleare. Sezione di Torino, \textit{I-10125}, Torino, Italy}

\author{M. Taoso}%\email{taoso@to.infn.it}
\affiliation{Istituto Nazionale di Fisica Nucleare. Sezione di Torino, \textit{I-10125}, Torino, Italy}

%%%%%%%%%%%%%%%

\date{\today}

\begin{abstract}
The Large Magellanic Cloud (LMC) is the biggest satellite galaxy of the Milky Way and provides a unique laboratory for high-energy astrophysics and dark matter physics.
In this work, we develop an end-to-end numerical description of the cosmic-ray (CR) transport and of the associated non-thermal emission in the LMC, extending the public \texttt{DRAGON} and \texttt{HERMES} codes.
Within this framework, we compute the diffuse synchrotron radiation of CR electrons in the LMC, and compare the predictions with observed low-frequency radio maps.
Because electron diffusion imprints a characteristic morphology on the radio signal, these comparisons enable to infer the effective average of the LMC diffusion coefficient.
We find 
$D_0=3-6\times 10^{28}\,{\rm cm^2/s}$ 
at 1 GeV, which is comparable to, but slightly larger than, values typically obtained for the Milky Way. 
More generally, this work provides a scalable tool for interpreting non‑thermal signals in nearby galaxies and constraining their CR transport.
\end{abstract}

\maketitle
\flushbottom

\section{Introduction} \label{sec:intro}

The Large Magellanic (or Milky) Cloud (LMC) is the most massive satellite galaxy of the Milky
Way. Its proximity and precisely determined distance~\cite{Pietrzynski:2019cuz} allows telescopes to study its properties and morphology in greater details than those of more distant galaxies. Owing to its relatively large mass and high level of astrophysical activity - hosting supernova remnants and star-forming regions - the LMC serves as an exceptional laboratory for high-energy astrophysics. 
LMC is a unique target in the sky and it is studied in connection with a variety of physical processes.
The focus of this work is on the transport of cosmic rays (CRs).

In the Milky Way, the inference of CR properties from the observation of their radiative diffuse emission, predominantly originating in the Galactic disk, is complicated by line of sight confusion.
Our external perspective of the LMC, combined with its modest inclination angle $\sim30{^\circ} $\cite{2001AJ....122.1807V} significantly mitigates this issue.

Analyses of LMC supernovae remnants~\cite{Bozzetto:2017,Neronov:2017syf} and diffuse $\gamma$-ray emission~\cite{Abdo:2010pq,Murphy:2012,Fermi-LAT:2015bpm,Foreman:2015gqa,Tang:2017xcz} suggest that the transport of CRs in the LMC can be described as diffusive propagation in a turbulent medium.
In this study, we develop a full numerical solution of the CR transport equation in the LMC and compute the synchrotron emission associated to CR electrons.
We employ the \texttt{DRAGON}~\cite{Evoli:2016xgn} and \texttt{HERMES}~\cite{Dundovic:2021ryb} codes. They are originally designed to describe CR propagation in the Galaxy. We extended them to describe a galaxy different from the Milky-Way (in \texttt{DRAGON}) and to consider an external observer's viewpoint (in \texttt{HERMES}).
These improvements enable us to accurately predict the LMC diffuse emission at low radio frequencies, arising from CR synchrotron radiation.

By comparing these predictions with observations from the Murchison Widefield Array (MWA) telescope~\cite{2018MNRAS.480.2743F}, we constrain the description of CR transport in the LMC.
In particular, the LMC's nearly face-on orientation allows us to tightly constrain the degree to which the spatial distribution of CR electrons is broadened by diffusion relative to their initial, localized injection around CR accelerators.

The paper thus provides a measurement of the effective average of the diffusion coefficient in the LMC. 
For the first time, the diffusion coefficient of the LMC is inferred from the morphology of the low-frequency radio emission.

Understanding diffusion in the LMC is not only relevant in the context of CR physics, but also a crucial ingredient for deriving particle dark matter (DM) bounds from radiative emission of DM annihilation or decay products.
Due to the substantial DM content and proximity to Earth, LMC is in fact considered as one of the most promising targets for DM indirect searches~\cite{Regis:2021glv}.

\section{Data and Methodology} 
\label{sec:data}
\begin{figure}
    \includegraphics[width=\columnwidth]{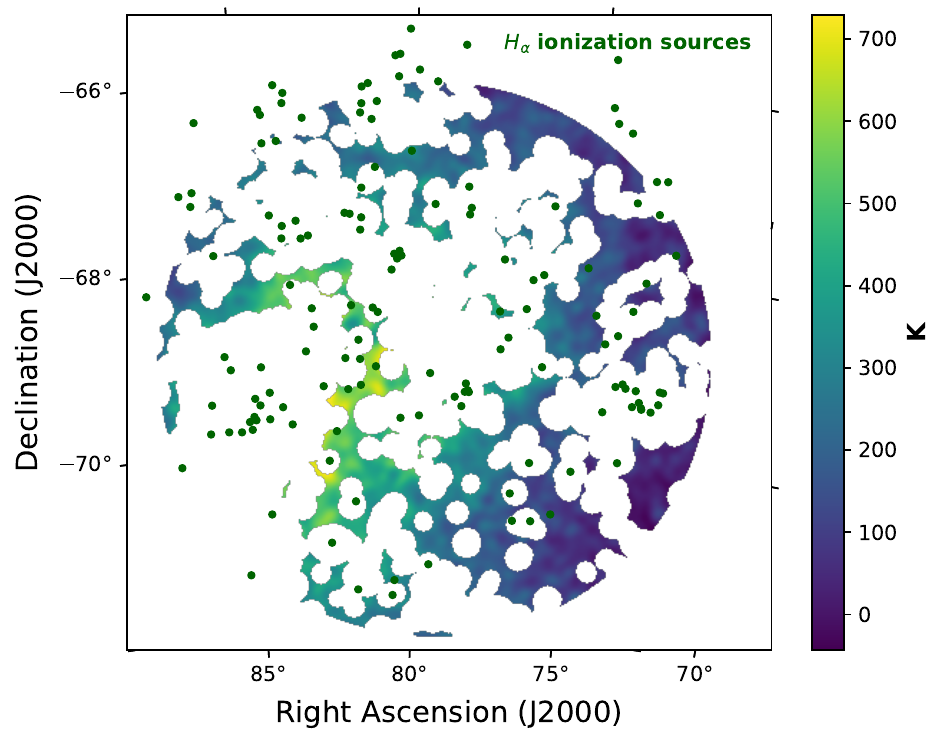}
    \caption{MWA radio map at 88 MHz~\cite{2018MNRAS.480.2743F}, expressed in brightness temperature (in K), along with the associated segmentation mask. The green dots correspond to the $H_{\alpha}$ regions from Ref.~\cite{2012ApJ...755...40P}.}
    \label{fig:Data_map_88}
\end{figure}

\textbf{Data}.
In our analysis, we make use of low-frequency MWA radio continuum maps, obtained from mosaics of the GaLactic Extragalactic All-Sky MWA (GLEAM) survey, and presented in Ref.~\cite{2018MNRAS.480.2743F}.
The maps span four frequency ranges, 88 MHz (72-103 MHz), 118 MHz (103-134 MHz), 155 MHz (139-170 MHz), and 200 MHz (170-231 MHz). The FWHM of the angular beam is $5.9'\times 5.0'$, $4.3'\times 3.7'$, $3.3'\times 2.8'$ and $2.7'\times 2.3'$ going from the lowest to the highest frequency bin. The RMS of the maps is around a few mJy/beam depending on the position in the sky and on the frequency.
The choice of these maps has two reasons.
First, at these frequencies we can assume the diffuse emission in the LMC to be dominated by synchrotron radiation, with other components, like free-free emission, being subdominant, and not included in our model.
Second, since we look for a diffuse emission and thus aim to probe large angular scales, we need interferometric images that are not affected by missing flux on scales smaller than the LMC size.
The shortest MWA baseline for the GLEAM observations is 7.7 m, which translates into a largest angular scale available of $29^\circ$ at 76 MHz and $10^\circ$ at 227 MHz~\cite{2018MNRAS.480.2743F}.
We derive the rms noise maps and identify bright sources using the software SExtractor~\cite{Bertin:1996fj}, and following the approach described in Ref.~\cite{Regis:2014koa}.
The small-scale discrete sources identified by SExtractor are masked, again because we are interested in studying the diffuse emission. Moreover, we smoothed all the maps with a Gaussian circular beam of $4'$ in order to smear additional small-scale fluctuations out.
As an example, the radio map  at a central frequency of 88 MHz with its corresponding segmentation mask is shown in Fig.~\ref{fig:Data_map_88}.

\textbf{Source distribution}.
Observational data from non-thermal radio emission~\cite{Hughes:2006me,2018MNRAS.480.2743F}, supernova remnants~\cite{Bozzetto:2017,Neronov:2017syf}, and $\gamma$-rays \cite{Abdo:2010pq,TheFermi-LAT:2015lxa} suggest that the CR distribution of the LMC is characterized by intermediate-scale and small-scale inhomogeneities (on top of a smooth diffuse sea), due to the presence of a variety of active regions that inject freshly accelerated particles in the medium. In this work, we aim at modeling this ``patchy'' component. 

In particular, we assume 
that the diffuse CR electrons are produced by a population of localized sources
which trace the ionized $H_{\rm{II}}$ regions from the catalog in Ref.~\cite{2012ApJ...755...40P}. Our approach is similar to the one adopted in Ref.~\cite{CherenkovTelescopeArray:2023aqu}.
We include only regions with $H_{\alpha}$ luminosities $L_{H_{\alpha}}$ above $10^{37} \; \rm{erg}/s$ in order to remain consistent with the assumption of steady injection. We then convert $H_{\alpha}$ luminosities into ionizing luminosities $L_{\rm{ion}}$ following the relation
\begin{equation}
    L_{\rm{ion}} = L_{H_{\alpha}} /2.2(1-f_{\rm{esc}}),
\end{equation}
where $f_{\rm{esc}}$ is the escape fraction of Lyman continuum photons from individual $H_{\rm{II}}$ regions, taken again from Ref.~\cite{2012ApJ...755...40P}.  

For the numerical implementation, we adopt a source distribution $Q(\vec{r},E)$ of the form
\begin{equation}
     Q(\vec{r},E) = \sum_k\,S_k(E)\, \frac{\exp{ [(\vec{r} - \vec{r_{k}})^2/\sigma_{\rm{src}}^2}]}{\sqrt{(2 \pi)^3} \sigma_{\rm{src}}^3} ,
    \label{eq:source_distribution}
\end{equation}
where $\vec{r}_k$ is the position of the ionized $H_{\rm{II}}$ region $k$, $\sigma_{\rm{src}}$ is its spatial extension, and $S_k(E)$ is the energy spectrum. 
The luminosity of each region is $L_k=a_k\,L$, where $a_k$ is the fractional contribution proportional to the ionizing luminosity described above and $L$ is
the total LMC luminosity. We take $S_k(E)$ to be a power-law, as detailed in the Supplementary Material (SM), with the same spectral index for all regions, and with normalization given by the luminosity through $L_k=\int dE\,E\,S_k(E)$. 
The total luminosity $L$ is taken as a free parameter in our analysis. As a reference value, a luminosity of
$L_{\rm ref}=6.5\times 10^{37} \; \rm{erg}/s$ 
is obtained assuming that CRs are injected by a population of supernovae (SN) with each event releasing an energy of $10^{51} \; \rm{erg}$, of which a fraction of $10^{-3}$ goes into CR electrons, and considering a rate of explosion of $0.002 \; \rm{SN}/\rm{yr}.$
Concerning the spatial extension, we adopt a value of  $\sigma_{\rm{src}}=40\; \rm{pc},$ which is well below the resolution of the spatial grid used to solve numerically the CR transport equation. In practice, the source distribution resembles a collection of point sources.

\textbf{Cosmic-ray propagation}.
The number density of CR electrons is obtained as the steady-state solution of the CR transport equation, sourced by the previously described distribution.

We choose to consider a minimal setup where the transport is only driven by spatial diffusion and energy losses.
We do not include advection and momentum diffusion in our treatment.  
A similar approach was used in the past in the first three-dimensional models of the Milky Way, see for instance \cite{Gaggero:2013rya}, and in the recent description of the gamma-ray emission from the LMC from Ref.~\cite{CherenkovTelescopeArray:2023aqu}.

As far as diffusion is considered, we treat it as isotropic with a coefficient of the form
\begin{equation}
    D(p,r,z) = D_0 \left( \frac{p}{p_0}\right)^{\alpha_D} e^{r/r_{\rm{scale}}} e^{z/z_h},
\label{eq:Diffcoeff}
\end{equation}
where $D_0$ sets the normalization, $p_0$ is a reference momentum that we take $1 \; \rm{GeV}$, $\alpha_D$ parametrizes the momentum dependence, $r$ is the radial coordinate in the LMC plane and $z$ in the perpendicular direction.
The scales $r_{\rm{scale}}$ and $z_h$ parameterize the extension of the diffusive region. 
The transport equation is solved using the numerical code \texttt{DRAGON}.
Since this framework is optimized for the Milky Way, we modified it to model the LMC, adopting appropriate models for the magnetic field, $H_I$ gas distribution and interstellar radiation field. 
A particularly significant modification of the \texttt{DRAGON} code regards the orientation: given the relative inclination between the Milky Way and the LMC, the computation is performed in a rotated frame. All the modifications are detailed in the SM.

\textbf{Synchrotron maps}. 
The synchrotron emission produced by the electron distribution is computed using the \texttt{HERMES} code \cite{Dundovic:2021ryb}. 

Also in this step of the computation, the \texttt{HERMES} framework was significantly modified because the code structure is originally designed to consider an observer that is located {\it inside} the Milky Way and observes the emission ``edge-on''. Here, the observer is located far away from the region that corresponds to the \texttt{DRAGON} propagation box and the code was therefore adapted in order to consider a distant galaxy and to take into account only the relevant portion of the line of sight. Again, more details about the relevant modifications to the numerical packages are presented in the SM.

The resulting emission maps, at a given frequency $\nu$, are provided in terms of the brightness temperature $T_b:$
\begin{equation}
    T_b(\nu,l,b) = \frac{c^2}{2\nu^2 k_B} I_b(\nu,l,b), 
    \label{eq:brightness_T}
\end{equation}

where $I_b$ the synchrotron intensity at the galactic coordinates $l,b$.

\begin{figure}
    \includegraphics[width=\columnwidth]{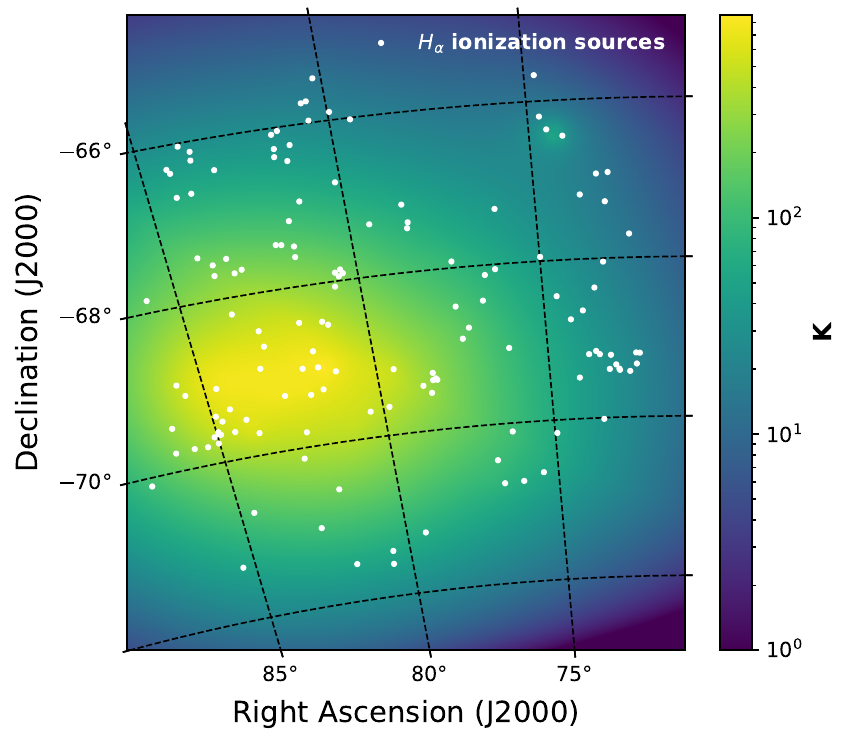}
    \caption{
    Synchrotron emission at 88 MHz, expressed in brightness temperature (in K), for the best-fit model of the joint analysis. The white dots correspond to the $H_{\alpha}$ regions from Ref.~\cite{2012ApJ...755...40P}.
    }
    \label{fig:Benchmark}
\end{figure}

\section{Analysis and Results}
\label{sec:res}

\begin{figure}
    \includegraphics[width=\columnwidth]{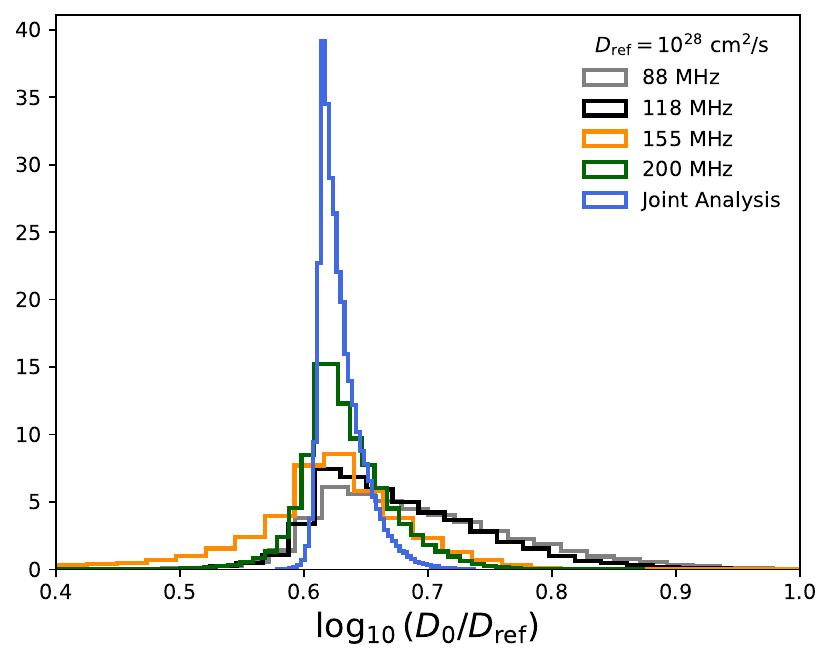}
    \caption{
    Posterior distributions of $D_0$ for the analyses of the individual observational maps and the combined analysis.
    }
    \label{fig:Post_benchmark_Do}
\end{figure}

\begin{figure}
    \includegraphics[width=\columnwidth]{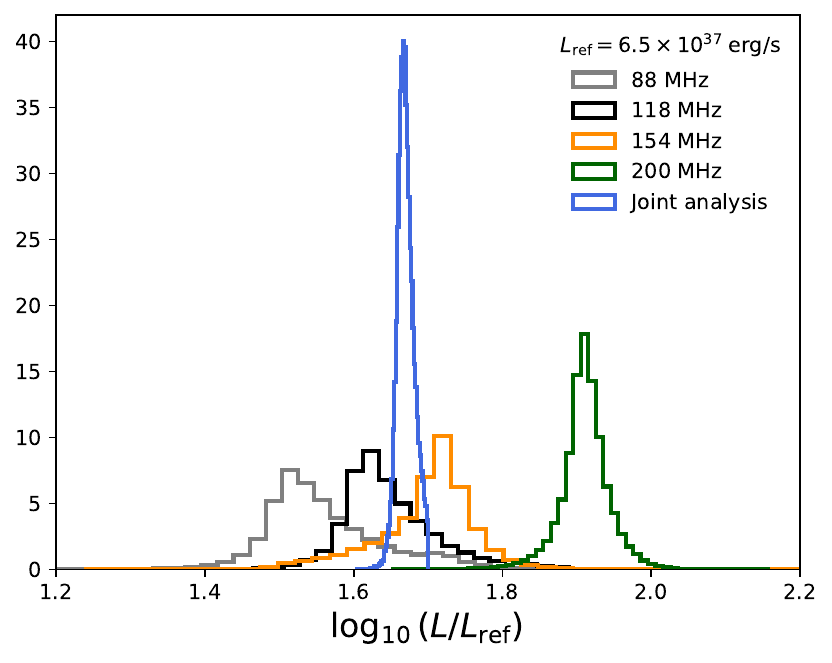}
    \caption{
    Posterior distributions of $L$ for the analyses of the individual observational maps and the combined analysis.
    }
    \label{fig:Post_benchmark_Lo}
\end{figure}

We analyze a circular region of interest (ROI) with a radius of 3 degrees centered at RA=79.4 and DEC=-69.03, which corresponds to the center of the stellar distribution. 
For each radio map, we implement a likelihood of the form $\mathcal{L} = e^{-\chi^2/2}$
where the $\chi^2$ is given by 
\begin{equation}
    \chi^2 = \frac{1}{N^{\rm{FWHM}}_{\rm{pix}}} \sum_{i=1}^{N_{\rm{pix}}} \left( \frac{T^i_{\rm{model}} - T^i_{\rm{obs}}}{\sigma^{i}_{\rm{rms}}} \right)^2,
    \label{eq:chi_square}
\end{equation}
with the sum running over the unmasked pixels $N_{\rm{pix}}$ inside the ROI. $N^{\rm{FWHM}}_{\rm{pix}}$ is the number of pixels in an angular beam, $T^i_{\rm{obs}}$ is the brightness temperature of the radio map in a given pixel and $\sigma^{i}_{\rm{rms}}$ the rms noise.
The brightness temperature predicted by our model, $T^i_{\rm{model}},$ is the sum of a constant term and the synchrotron emission from the electrons population described before. The former term is introduced to account for a potential zero-level offset of the map and a possible large scale emission, and it is described by a free parameter for each map. 
Concerning the latter term, we adopt a benchmark model where all relevant parameters are fixed as in Table~\ref{table:params} except for $D_0$ and $L$ which are kept as free parameters. 
We explore the parameter space defined by these two quantities and the constant terms using the Markov Chain Monte Carlo sampler Emcee~\cite{2013PASP..125..306F}.
The analysis is performed independently for each radio map as well as combining (multiplying) the likelihoods for the four radio maps in a joint fit (6 scanned parameters).
The posterior distributions for $D_0$ and $L$ are shown in Figs.~\ref{fig:Post_benchmark_Do} and ~\ref{fig:Post_benchmark_Lo}. We find that the distributions obtained from the analyses of the individual maps are in good agreement with each other.
From the joint analysis, we find $D_0=4.22^{+0.21}_{-0.10}\times 10^{28}\,{\rm cm^2/s},$ where the central value corresponds to the posterior median and the interval spans the 16th to 84th percentiles.
In Table~\ref{table:best-fit}, we report the results for $D_0$ for each observational map, along with the $\chi^2$ per degree of freedom, $\chi^2/{\rm dof}$, where the degrees of freedom are obtained from the number of independent angular beams, i.e. $N_{\rm pix}/N_{\rm pix}^{\rm FWHM}$. The values of $\chi^2/{\rm dof}$ show our fits are reasonable.
The emission corresponding to the best-fit model is shown in Fig.~\ref{fig:Benchmark}, for the representative case of the 88 MHz map.

\begin{table}[h!]
  \centering
  \caption{Median values of $D_0$ and $\chi^2/{\rm dof}.$}
  \begin{tabular}{|c|c|c|} 
    \hline
    \textbf{Radio map} & \rule{0pt}{2.6ex} $D_0$ [$10^{28}$ cm$^2$/s]  & $\chi^2/{\rm dof}$\\ 
    \hline
    88 MHz & $4.78_{-0.63}^{+1.17}$ & 0.97\\ 
    118 MHz & $4.62_{-0.50}^{+0.94}$ &  1.07 \\
    154 MHz & $4.18_{-0.51}^{+0.55}$ &  1.68 \\
    200 MHz & $4.25_{-0.22}^{+0.39}$ &  2.07 \\
    \hline
  \end{tabular}
\label{table:best-fit}
\end{table}

\section{Discussion and Conclusions} \label{sec:conc}
While the posterior distributions of $D_0$ at different frequencies are remarkably overlapping, the ones of $L$ show larger differences. Even though they are not incompatible with each other, they seem to follow a increasing trend with frequency suggesting a slightly different description of the electron spectrum of injection.
More importantly, the preferred values are significantly larger than our reference value $L_{\rm ref}$.
This might be due to a number of reasons including: larger SN injection luminosity or rate~\cite{Bozzetto:2017}; a larger fraction of energy going into CR electrons; a larger magnetic field (as possibly suggested by some recent works~\cite{2022MNRAS.510...11H,Seta:2022uoy}); a different spectral index for the injection spectrum of electrons, in particular below 1 GeV that is a regime we do not probe in our analysis.
All these modifications change the overall synchrotron intensity but do not significantly affect the results about $D_0$, as shown in the SM. Indeed, the main point of our work is that since LMC is nearly face-on, the level of patchiness of the synchrotron emission can strongly constrain the degree to which the spatial distribution of CR electrons is broadened by diffusion relative to the injection morphology localized around CR sources. This is uncorrelated to an overall rescaling of the signal, and allows us to greatly narrow down systematic uncertainties on $D_0$, despite remaining large for other ingredients of the computation.
The LMC synchrotron emission around 100 MHz is mostly generated by CR electrons at GeV energies. At such low energies, the electrons travel in a diffusive regime, where the residency time $z_h^2/D\sim 10$ Myr is much shorter than the cooling time associated to energy losses $\tau_c\simeq 100$ Myr. This makes our result robust against variations in the characterization of the interstellar medium and of the associated radiative losses, as we explicitly tested.
In our description, the only quantity other than $D_0$ that can have a relevant impact on the patchiness of the electron distribution is the vertical size $z_h$ of the diffusion box. 
Essentially, we constrain the ratio $z_h/D_0\simeq 10\, {\rm Myr}/1.5\, {\rm kpc}$, for GeV electrons.
For reasonable scenarios, the impact of the systematic uncertainty of $z_h$ on the determination of $D_0$ is $\mathcal{O}(1)$ and our work implies $D_0=3-6\times 10^{28}\,{\rm cm^2/s}$ at 1 GeV.
This is shown the SM, where we report a few tests with different modeling, to show the robustness of the determination of $D_0$, and two example maps where the link between the patchiness of the radio image and the diffusion coefficient can be visualized.

As described above, our derivation is based on a model where, on top of a large-scale CR population uniformly distributed across the LMC disk arising from CRs accumulated over long time scales and accounted for through the spatially uniform term in the fit, there are localized sources from energetic star forming regions that provide the dominant CR injection at small scales.
This description is supported also by Fermi-LAT LMC observations of $\gamma$-ray emission from CRs~\cite{Abdo:2010pq,TheFermi-LAT:2015lxa}, where a large-scale component plus three to four degree-scale regions were found, by the study of radio sources in the LMC, which show strong correlation with star forming regions~\cite{Hughes:2006me,2018MNRAS.480.2743F} and by supernova remnant energetics~\cite{Bozzetto:2017,Neronov:2017syf}.

The values of $D_0$ preferred by our analysis is similar to the one of our Galaxy, even though around a factor of two larger with respect to what found in studies of the Milky-Way with pure-diffusion models~\cite{Bueno:2022bdc}.

Note that the diffuse radio-emission modeling developed here provides a useful framework to search for signals from DM annihilation or decay in the LMC. We plan to investigate this topic soon.

In conclusion, this work developed a numerical tool for studying non-thermal diffuse signals in nearby galaxies, and applied it to the LMC. Thanks to nearly face-on LMC orientation, the comparison between predicted synchrotron diffuse emission from CRs and observational radio maps allowed us to constrain the degree to which the spatial distribution of CR electrons is broadened by diffusion relative to their initial, localized injection around CR accelerators.
The paper thus presented a new method to measure the diffusion length in the LMC, and determine a narrow range for the LMC diffusion coefficient $D_0=3-6\times 10^{28}\,{\rm cm^2/s}$.

\acknowledgments
JRC acknowledges the support of the Natural Sciences and Engineering Research Council of Canada (NSERC), funding reference number RGPIN-2020-07138, and the NSERC Discovery Launch Supplement, DGECR-2020-00231. JRC acknowledges support from INFN through the Senior Research Fellowship 
program (Grant No. 27076). MR, MT and DG acknowledge support from the  Research grant TAsP (Theoretical Astroparticle Physics) funded by \textsc{infn}. DG further acknowledges support from the TEONGRAV grant funded by \textsc{infn}. The work of MR is supported by the Italian Ministry of University and Research (MUR) via the PRIN 2022 Project No. 20228WHTYC – CUP: D53C24003550006. MT acknowledges the research grant “Addressing systematic uncertainties in searches for dark matter No. 2022F2843” funded by MIUR. 

\bibliographystyle{apsrev4-2}
\bibliography{biblio}

\appendix
\section{Ingredients of the transport equation}
We consider a propagation setup based on isotropic diffusion and momentum losses. The steady-state equation providing the equilibrium distribution $N(\vec r,p)$ of CR electrons as a function of their position $ \vec r$ in the LMC and their momentum $p$ is written as follows:
\begin{equation}
\nabla\cdot\big(D(\vec r,p)\,\nabla N(\vec r,p)\big) + 
\frac{\partial}{\partial p}\!\big(\dot{p}\,N(\vec r,p)\big)
+ Q(\vec r,p) = 0 \, ,
\end{equation}
with the usual notation where $D$ is the diffusion coefficient, $\dot{p}$ is the momentum loss term, and $Q$ is the injection term.
In the following, we provide details about the description of these ingredients and of the numerical solution of the equation.

\subsection{Source Distribution}
As mentioned in the main text, we model cosmic-ray (CR) sources from the catalog in Ref.~\cite{2012ApJ...755...40P}.
The spatial distribution in Eq.~\ref{eq:source_distribution} is obtained from the sky coordinates of the sources, and by assuming they all lie on the LMC galactic plane.

The spectral form of the CRs injection is taken to be:
\begin{equation}
    \frac{dP_{\rm{inj}}}{dE} = \beta \left( \frac{E}{E_0}\right)^{-\alpha_{\rm{inj}}} \exp{\left[ -E/E_{\rm{cut}} \right]},
    \label{eq:injection_spectrum}
\end{equation}
where $\beta$ is the Lorentz factor, $\alpha_{\rm{inj}}$ is the injection spectral index, $E_0$ a pivot energy and $E_{\rm{cut}}$ is the cut-off energy, and we have adopted values reported in Ref.~\cite{CherenkovTelescopeArray:2023aqu}, as summarized in Table~\ref{table:params}. The CR injection luminosity of each individual source is 
\begin{equation}
    L_k = A_k\int_{E_{\rm{min}}}^{E_{\rm{max}}} dE \, E \,\frac{dP_{\rm{inj}}}{dE}\;.
    \label{eq:norm_lum}
\end{equation}
In the main text, we introduced the energy spectrum $S_k(E)$ which is given by $S_k(E)=A_k\,dP_{\rm{inj}}/dE$.
The normalization factor $A_k$ is computed as
\begin{equation}
    A_k = \frac{a_k L}{\int_{E_{\rm{min}}}^{E_{\rm{max}}} dE \; E\; dP_{\rm{inj}}/dE }.
    \label{eq:norm_lum}
\end{equation}
where $a_k$ is the fractional contribution to the total luminosity, obtained from the ionization luminosity reported in the catalog of Ref.~\cite{2012ApJ...755...40P}.

\begin{table}[h!]
  \centering
  \caption{Parameters of the benchmark model.}
  \begin{tabular}{cc} 
    \toprule
    \textbf{Parameters} & \textbf{Value} \\ 
    \midrule
    $x_{\rm{min}}$, $x_{\rm{max}}$ & -3 kpc, 6 kpc \\ 
    $y_{\rm{min}}$, $y_{\rm{max}}$ & -4 kpc, 3 kpc \\ 
    $z_{\rm{min}}$, $z_{\rm{max}}$ & -5 kpc, 5 kpc \\
    $\sigma_{\rm{src}}$ &  40  pc\\
    $\rm{Dim}_x$ & 90 points \\ 
    $\rm{Dim}_y$ & 70 points \\ 
    $\rm{Dim}_z$ & 100 points \\ 
    $E_{\rm{min}}$, $E_{\rm{max}}$ & [0.1,$10^4$] GeV \\ 
    $D_{\rm{ref}}$ & $10^{28}\rm{cm}^2/\rm{s}$ \\ 
    $r_{\rm{scale}}$, $z_h$ & [5,1.5] kpc \\ 
    $\alpha_D$ & 1/3 \\ 
    $p_0$ & 1 GeV \\ 
    $B_0$ & 4.3 $\mu G$ \\ 
    $L_{\rm{ref}}$ & 6.5 $\times 10^{37}$ erg/s \\
    $E_0$ & 0.1\,GeV \\
    $\alpha_{\rm{inj}}$ & 2.65 \\ 
    $E_{\rm{cut}}$ & $10^{5}$ GeV \\ 
    \bottomrule
  \end{tabular}
\label{table:params}
\end{table}

\subsection{Energy Loss Terms}
\label{sec:energy_loss}
At the energies relevant to our study, the dominant timescale is the diffusion one (similarly to the case of the Milky Way, see for instance \cite{Evoli:2016xgn}). The role of energy losses is sub-dominant but non-negligible and we model them in a consistent way with {\tt DRAGON}. The main processes that need to be considered are bremsstrahlung, synchrotron, and inverse Compton emissions.
In order to model them, we need to specify the spatial dependence of the H\textsc{i} gas, magnetic field and the Interstellar Radiation Field (ISRF).

\textbf{Gas density}. The main component considered in this work is neutral atomic hydrogen. We neglect molecular hydrogen and ionized gas because they are subdominant components in the LMC~\cite{Regis:2021glv}. We assume an H\textsc{i} gas density profile of the form \cite{Bustard_2020}:
\begin{equation}
    n_{\rm{HI}}(r,z) = \rho_{H,0} \; \rm{sech}\left( \frac{r}{a_g}\right) \rm{sech}\left( \frac{|z |}{b_g} \right),
\end{equation}
where the coordinates $r$ and $z$ correspond to the cylindrical coordinates in the LMC frame, while $a_g = 1.7$ kpc and $b_g=0.34$ kpc are the length and height scales, respectively. $\rho_{H,0}$ is the central density, which has been derived in Ref.~\cite{Regis:2021glv} from H\textsc{i} rotation curves~\cite{Kim:1998}, finding a value of $\rho_{H,0} = 0.8 \; \rm{cm}^{-3}$.

\textbf{Magnetic Field}. As for the spatial dependence of the diffusion coefficient, we model of the magnetic field strength with a double exponential form:
\begin{equation}
B_{\rm{mag}}(r,z) = B_0 \exp{(-r/r_{\rm{scale}})}\exp{(-|z|/z_h)},
\label{eq:bfield_mag}
\end{equation}
with $B_0$, $r_{\rm{scale}}$ and $z_h$ reported in Table~\ref{table:params}. 
In particular, we assume that the region where CR diffusive confinement is efficient correlates with the intensity of the magnetic field, therefore we take the same scales $r_{\rm{scale}}$ and $z_h$ in Eq.~\ref{eq:Diffcoeff} and Eq.~\ref{eq:bfield_mag}.
Concerning the magnetic field strength, below
we show a test using a different value and its impact on the results.
For what concerns the orientation of the magnetic field vector (which is relevant for the computation of the synchrotron emission along our line of sight), we consider a toroidal axisymmetric vector field lying on the plane of the LMC disk,
$\vec{B}(r,z) = B_{\rm{mag}}(r,z)\hat{\phi}$, with $\hat{\phi}$ describing the azimuthal direction. We tested that including a component perpendicular to the LMC disk (with a fixed ratio between parallel and perpendicular components) does reduce the emission, but this variation is re-absorbed in an enhancement of the $L_0$ parameter, and the impact on $D_0$ is negligible.

\textbf{ISRF}. We adopt a spatially uniform model for the ISRF, and the spectral energy density $u$ (energy per unit volume per unit frequency) is decomposed as a sum of Planck distributions $u_{P}$, following Ref.~\cite{CherenkovTelescopeArray:2023aqu}:  
\begin{equation}
    u(\nu) = \sum_{k=1}^5 \frac{U_k }{a T_k^4} u_{P}(\nu,T_k),
\end{equation}
where $T_k = \{2.73,35,330,3800,35000 \}$ K, $U_k = \{0.26,0.12,0.025,0.30,1.20 \}$  $\rm{eV}/\rm{cm}^3$, and $a=4\,\sigma/c$ with $\sigma$ being the Stefan-Boltzmann constant.

\begin{figure}
    \centering
    \includegraphics[width=1.2\columnwidth]
    {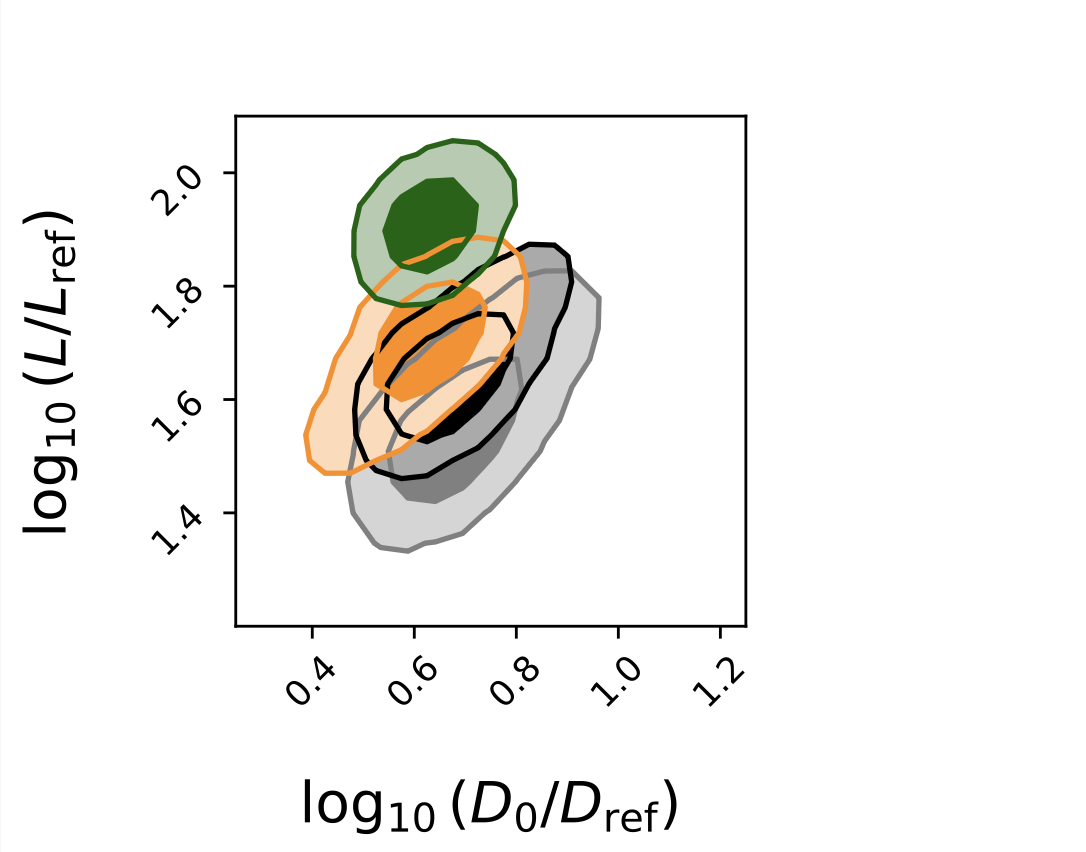}
    \caption{
    Contours at 68\% and 95\% credible level} 
    of $D_0$ and $L$ for the analyses of 
    the individual observational maps.
    \label{fig:corner_all_freq}
\end{figure}

\begin{figure*}[t]
    \centering
    \includegraphics[width=\textwidth]{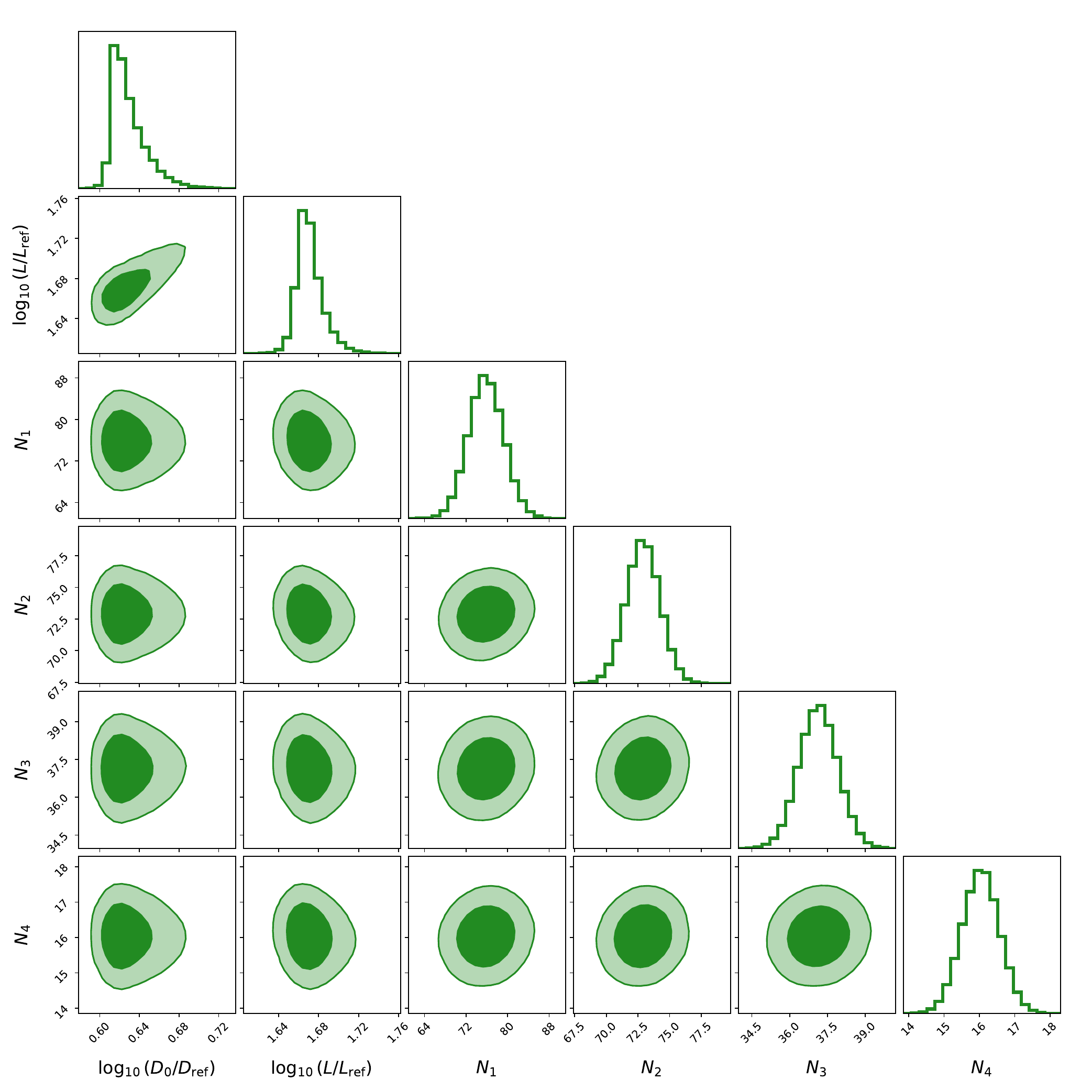}
    \caption{
    Posterior distributions of all the parameters for the combined analysis of 
    the observational maps.
    $N_i,$ $i=1,2,3,4$ correspond to the brightness temperature (in K) of the spatially uniform terms introduced in our model, see the main text for more details. 
    }
    \label{fig:corner}
\end{figure*}

\section{Numerical Implementation}
\label{sec:num_implementation}

The solution of the transport equation 
for the CR electron number density
is computed numerically through the \texttt{DRAGON} code~\cite{Evoli:2016xgn}. Such framework has been extensively used to compute transport processes in the Milky Way, for which is highly optimized. We modified several functions of the code, in order to describe the LMC source distribution, gas distribution, magnetic field and ISRF, as discussed in the previous Section. Modifying \texttt{DRAGON} with new ingredients for a galaxy different than the Milky-Way is simple, provided they are specified in the coordinates of the galaxy we are describing. However, such implementation is not yet compatible with the line-of-sight (LOS) integrator \texttt{HERMES}, which coordinate frame is fixed at the observer (i.e., in heliocentric Galactic coordinates). 
To overcome such limitation, the calculations within \texttt{DRAGON} are performed in a rotated frame with axes aligned with the Galactic coordinate system:
\begin{equation}
    \vec{r}_{\rm{DRAGON}} = R \; \vec{r}_{\rm{LMC}},
    \label{eq:rot}
\end{equation}
where $R$ is a rotation matrix that can be defined by knowing the inclination $i$ and position $\theta$ angles of the plane of the LMC. This has been derived following the methodology and numerical values of Ref.~\cite{2001AJ....122.1807V}. 
As far as \texttt{HERMES} is concerned, the position in the heliocentric Galactic coordinate system is obtained by a simple translation $\vec{r}_{\rm{HERMES}} = \vec{r}_{\odot}^{\,\rm{LMC}} +\vec{r}_{\rm{DRAGON}},$ with $\vec{r}_{\odot}^{\,\rm{LMC}}$ being the vector from the observer to the center of LMC.

To properly account for the anisotropic injection source distribution resulting from the sum of individual sources, we run \texttt{DRAGON} with a 3D+1 grid, namely, with three spatial dimensions plus energy. 

To specify the spatial distribution of the sources, we start from their right ascension ($\alpha_k$) and declination ($\delta_k$), taken from the catalog of Ref.~\cite{2012ApJ...755...40P}. Then, to compute their distances $d_k$
from the observer, we follow Ref.~\cite{2001AJ....122.1807V}. Assuming that all sources lie on the galactic plane of the LMC one gets:
\begin{equation}
    d_k = \frac{d_0 \cos{i}}{\cos{i}\cos{\rho_k} - \sin{i}\sin{\rho_k}\sin{(\phi_k - \theta)}},
\label{eq:source_distances}
\end{equation}
where $d_0$ is the distance from the observer to the center of the LMC, $i$ and $\theta$ are the inclination and position angles of the LMC plane~\cite{2001AJ....122.1807V}.
The angular coordinates $\rho_k$ and $\phi_i$ define the position of the source in the sky with respect to the center of the LMC.
They are computed from the sources coordinates $\alpha_k$ and $\delta_k$, again following equations in Ref.~\cite{2001AJ....122.1807V}.
Once the quantity $d_k$ is obtained, it is straightforward to compute the distance from the source to the center of the LMC.

All other ingredients in our LMC model are specified in LMC cylindrical coordinates and then rotated using the matrix $R$ introduced above.

\texttt{DRAGON} solves the transport equation on a numerical grid. Since the LMC  plane is rotated with respect to the Galacto-centric coordinate system, we adopt unequal minimum and maximum values for the grid boundaries. These limits are selected such that the 3D box contains all sources and the diffusion cylinder defined by $r_{\rm{scale}}$ and $z_h$.

The number of points in each axis of the grid is chosen to obtain a spatial resolution of $\sim 0.1 \; \rm{kpc}$. This translates to an angular resolution of $\theta\sim0.114$ degrees
which is approximately (slightly larger) the resolution of the $88 \, \rm{MHz}$ map and of our smoothing of the radio images, as described in the main text. 
The spatial extension of the sources is set to be $\sigma_{\rm{src}} = 40 \, \rm{pc}$.
Any choice $\sigma_{\rm{src}}\ll 0.1$ kpc has no impact on the results, as we explicitly checked. Thus, effectively, our assumption concerning the source sizes is that they are within 100 pc.
A list of the most relevant parameters in the configuration file of \texttt{DRAGON} for our benchmark model are reported in Table~\ref{table:params}.
A public available code containing the implementations for the LMC and that can be easily adapted to any nearby galaxy is currently under development.

\smallskip

The synchrotron surface brightness is given by 
\begin{equation}
    I_\nu (\nu,l,b) = \frac{1}{4\pi} \int_0^{\infty} ds\; A_s(\nu,\vec{r})\epsilon_\nu(\vec{r},\nu),    
\end{equation}
where $A_\nu$ is the absorption function and $\epsilon_\nu$ is the synchrotron emissivity, both detailed in Ref.~\cite{Dundovic:2021ryb}. The integration is performed along the LOS. There is no averaging over the telescope beam, since, as already mentioned, our spatial resolution is larger than it. Given the frequency range of interest, the effect of free-free absorption is not relevant in this context. 

The brightness temperature is then obtained from Eq.~\ref{eq:brightness_T}.
We perform the numerical calculations using
the public package \texttt{HERMES}, which numerically computes the emission from a population of electrons 
resulting from the numerical propagation in \texttt{DRAGON}. The code produces maps following the \texttt{HEALPIX} convention~\cite{Gorski:2004by} in the Galactic reference frame.
We choose a HEALPIX pixelization \texttt{nside} = 512, which matches the angular resolution mentioned above of our numerical analysis.

Note that a proper 3D description is essential. E.g., the emissivity $\epsilon_\nu$ depends on the magnetic field component $B_\perp$ that is perpendicular to the line of sight, and thus the exact orientation of $\vec B$ with respect to our line of sight is crucial to derive the synchrotron maps.

\section{Extended results of the analysis} \label{sec:extres}

In this section, we report additional details of the analysis presented in the main text.
The 2D posterior distributions of the two physical parameters considered in our statistical analysis, namely $D_0$ and $L$, as derived from each individual radio map, are shown in Fig.~\ref{fig:corner_all_freq}.
As already noticed in the main text, there is good agreement among different frequencies.
It is also important to note that the degree of degeneracy between $D_0$ and $L$ is very limited.

In Fig.~\ref{fig:corner}, we show the full corner plot for the combined analysis, which fits simultaneously all the radio maps. On top of $D_0$ and $L$, the figure reports the posterior distributions for the spatially uniform terms $N_i$ (one for each frequency $i$) included in our fit. We can appreciate that all parameters are well constrained and that there is no degeneracy between the $N_i$'s and $D_0$ or $L$.
We can see that the $N_i$ terms provide a small but non-negligible contribution by comparing the values in Fig.~\ref{fig:corner} with observational maps. E.g., at 88 MHz we have $N_1\simeq 75$ K which has to be compared with the values in Fig. 1 of the main text.

\begin{figure}[h!]
    \centering
    \includegraphics[width=\linewidth]{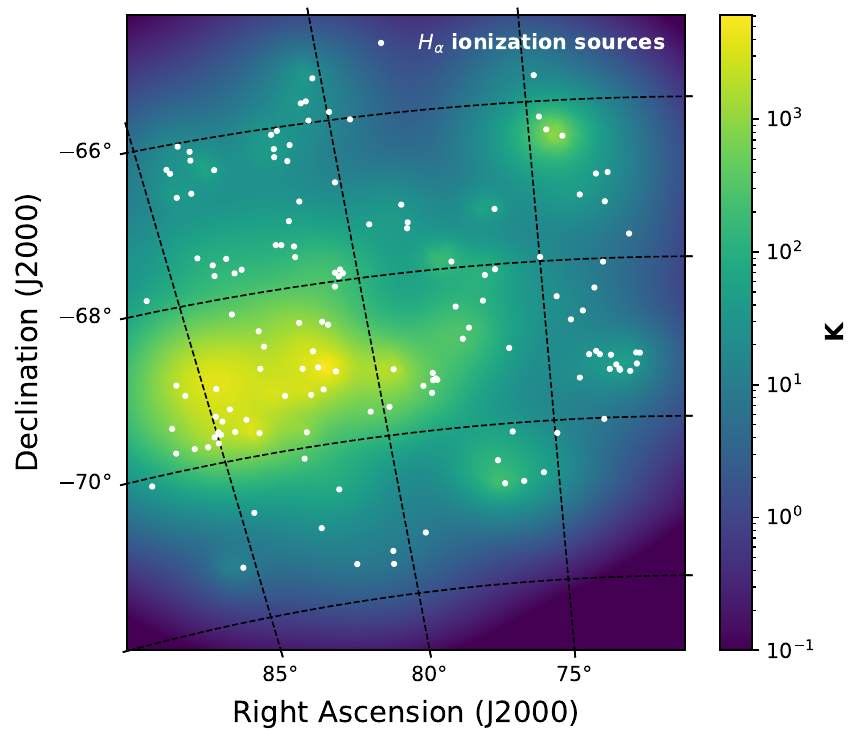}
    \caption{
    Same as in Fig.~\ref{fig:Benchmark} but for a much smaller diffusion coefficient: $D_0=10^{-2}\,D_{\rm ref}.$
    }
    \label{fig:Do_0.01}
\end{figure}

\begin{figure}[h!]
    \centering
    \includegraphics[width=\linewidth]{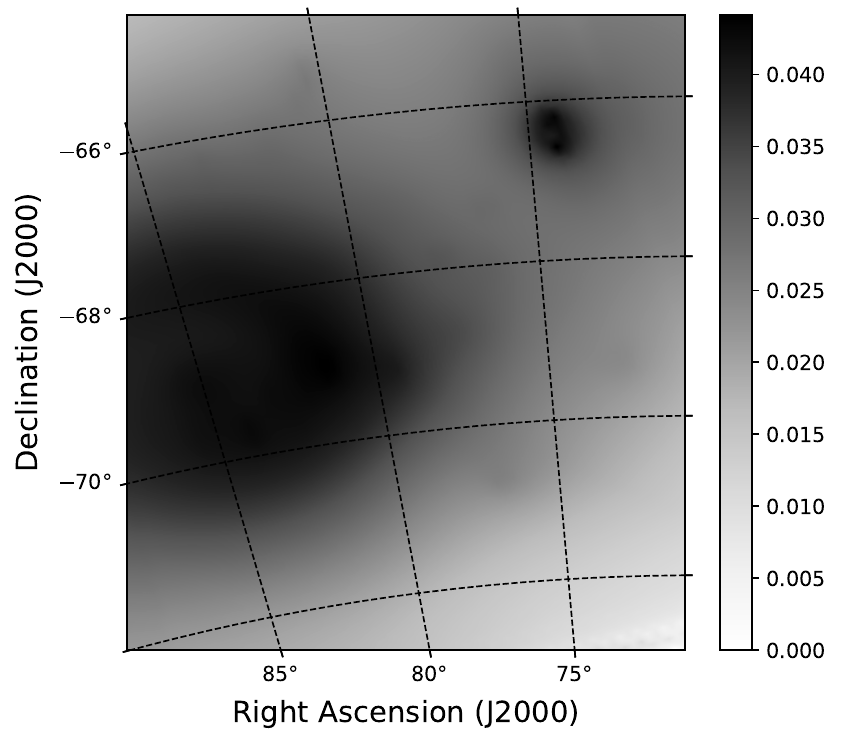}
    \caption{
    Relative difference of the synchrotron emission map (at 88 MHz) for the best-fit value of $D_0$ (as in Fig.~\ref{fig:Benchmark}) and for a much larger diffusion coefficient, $D_0=10^{2}\,D_{\rm ref}.$
    For sake of clarity, the two maps have been normalized to a common value in the outskirts.
    }
    \label{fig:Do_100}
\end{figure}

To understand better why $D_0$ does not suffer of significant degeneracy with other parameters, and, more in general, the reason why the current analysis is able to set stringent constraints on the diffusion coefficient, we show two examples in Figs.~\ref{fig:Do_0.01} and \ref{fig:Do_100}, where $D_0$ is taken to be very different from our best-fit case.
Fig.~\ref{fig:Do_0.01} assumes $D_0=10^{26}\,{\rm cm^2/s}$. It is clear that such a small value leads to a pattern of emission strongly localized around the individual sources. Observational maps show more diffuse patterns, which require sufficiently large diffusion lengths.
In Fig.~\ref{fig:Do_100}, we take $D_0=10^{30}\,{\rm cm^2/s}$, i.e., significantly larger than the best-fit value. In this case, the electrons escape more efficiently from the source regions leading to a more uniform emission.
The difference is less pronounced than in the case in Fig.~\ref{fig:Do_0.01}. 
To better visualize it, we normalize the map with larger diffusion coefficient so that in a pixel of the outskirt it coincides to the map of the best-fit $D_0$, and then in Fig.~\ref{fig:Do_100} we show their relative difference. This is clearly zero by construction in the outskirt, but highlights the lack of emission around the source region for the map with larger $D_0$. Observational maps require more patchiness and thus a lower $D_0$.

From these comparisons, we see that the diffusion coefficient has a non trivial impact on the morphological pattern of the synchrotron emission. 
Model uncertainties that affect the normalization of the maps, such as, e.g., the strength of the magnetic field, the spectrum of the electrons, or their luminosity, have little impact on the determination of $D_0$. Also, effects acting at large scales, such the ones associated to the parameters $N_i$, do not alter the pattern at intermediate scales that we discussed above, and this can explain the absence of degeneracy between $N_i$ and $D_0$.

\begin{figure}
    \includegraphics[width=\columnwidth]{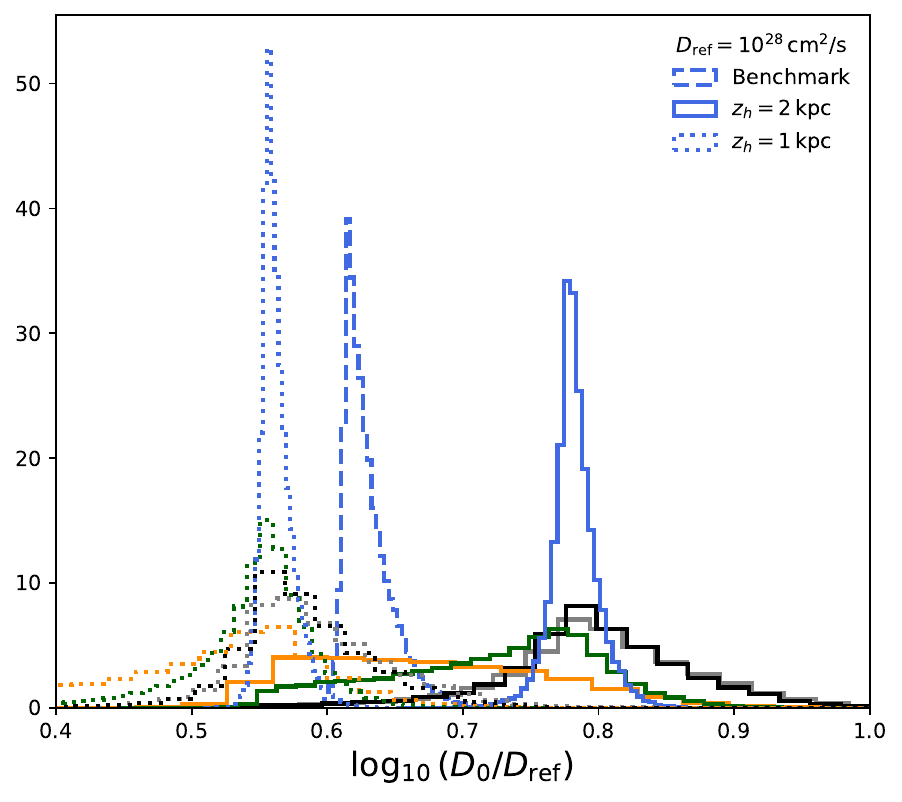}
    \caption{Posterior distributions of $D_0$ for the analysis of the individual observational maps and the combined analysis. The color coding is the same as in Fig.~\ref{fig:Post_benchmark_Do}.
    Solid and dotted lines are respectively for $z_h=2$ kpc and $z_h=1$ kpc. The dashed line shows the posterior distribution for the combined analysis of the benchmark model introduced the main text ($z_h=1.5$ kpc).
    }
    \label{fig:Do_test_zh_cases}
\end{figure}

\begin{figure}
    \includegraphics[width=\columnwidth]{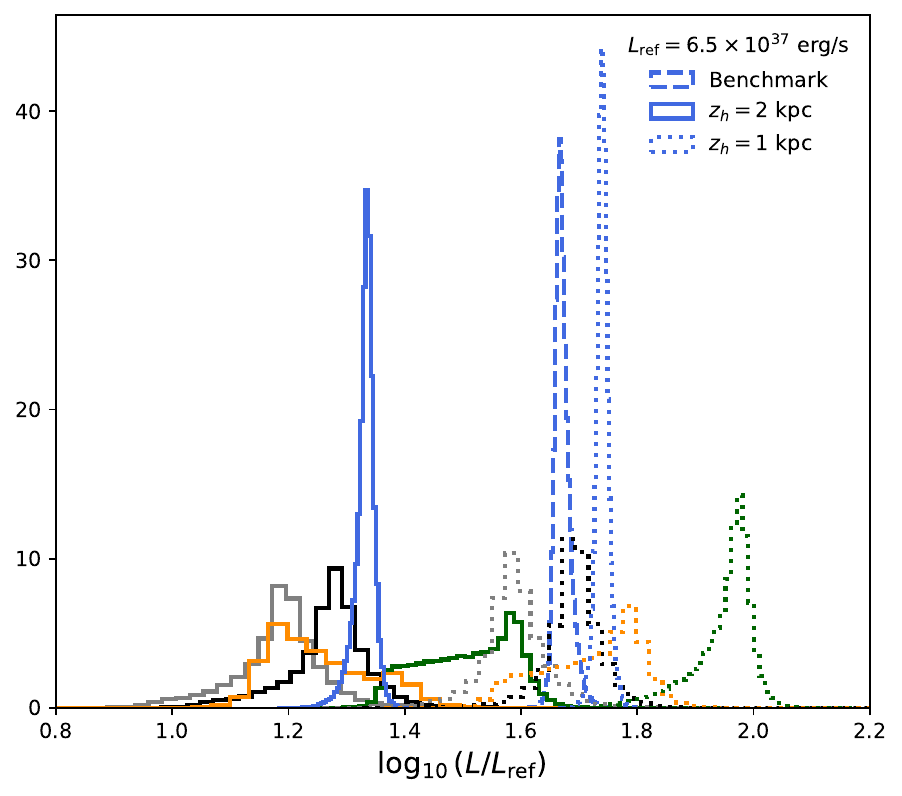}
    \caption{Same as in Fig.~\ref{fig:Do_test_zh_cases} but for $L.$
    }
    \label{fig:Lo_test_zh_cases}
\end{figure}

\subsection{Additional analyses}
\label{appendix:tests}

In this Section, we focus on the parameters of our LMC model that are most relevant for the synchrotron emission and analyses how variations in these quantities affect the inferred value of $D_0.$ Namely, with respect to our benchmark model considered in the main text and in the previous Section, we consider different choices for the size of the diffusive region, the magnetic field strength, and the electron energy spectrum injected by the sources. 

The diffusive region is parametrized by the scales $r_{\rm{scale}}$ and $z_h$ in Eq.~\ref{eq:Diffcoeff}. We consider a radial scale $r_{\rm{scale}}=5\,{\rm kpc}$, see Table~\ref{table:params}. This radius roughly encloses the LMC stellar distribution and corresponds to an angular radius of $\sim 5.7$ degrees, which is larger than the radius of our ROI. Thus variations of $r_{\rm{scale}}$ within reasonable intervals have negligible impact on our results.
The vertical extension $z_h$ is not well constrained. The nearly face-on orientation of the LMC hinders the determination of this quantity from the morphology of the radio emission. This method is instead applicable to edge-on galaxies --
for reference, scale heights of about 1–2 kpc have been inferred in Ref.~\cite{2018A&A...611A..72K} for a collection of galaxies of sizes comparable/larger than the LMC.
In our benchmark model, we have taken $z_h=1.5$ kpc and in the following we explore the cases of a smaller ($z_h=1$ kpc) and larger ($z_h=2$ kpc) diffusion height. The results are shown in Figs.~\ref{fig:Do_test_zh_cases} and~\ref{fig:Lo_test_zh_cases}.
Some degree of degeneracy between $z_h$ and $D_0$ is expected: for smaller values of $z_h$ CR electrons escape the diffusion region more efficiently (smaller $D$ at large $z$) and emit less synchrotron radiation far from the disk (smaller $B$ at large $z$), something that can be compensated by a smaller diffusion coefficient and an enhancement in the injected luminosity. The opposite trend is expected if $z_h$ increases.
These behaviors are indeed confirmed in Figs.~\ref{fig:Do_test_zh_cases} and~\ref{fig:Lo_test_zh_cases}. 
For $z_h=2$ kpc, we find a slight improvement in the value of $\chi^2/{\rm dof},$ and the median value of $D_0$ increases approximately linearly with $z_h.$ From the joint analysis, we find $D_0=6.04^{+0.23}_{-0.17}\times 10^{28}\,{\rm cm^2/s}$ (median value and 68\% credible interval).
Let us stress that, although the scale $z_h$ of LMC is not well known, values around 2-3 kpc have been typically found in galaxies
much larger than the LMC~\cite{2018A&A...611A..72K,2019A&A...632A..13S}.
For $z_h=1$ kpc, instead, we obtain a moderate worsening of $\chi^2/{\rm dof}$ and the shift in the $D_0$ distribution is smaller than that observed in the $z_h=2$ kpc case.
We obtain, for the joint analysis, $D_0=3.63^{+0.11}_{-0.06}\times 10^{28}\,{\rm cm^2/s}$.
In summary, for $z_h$ between 1 and 2 kpc, we constrain the diffusion coefficient in the range $D_0=3-6\times 10^{28}\,{\rm cm^2/s},$ as mentioned in the main text. 

In conclusion, these tests suggest that the uncertainty on $z_h$ are likely to influence the determination of $D_0$ by an $\mathcal{O}(1)$ factor.
This degeneracy means that the ratio $z_h/D_0$ remains approximately constant in our tests, as expected for GeV CR electrons in a diffusive regime, with $z_h/D_0\simeq 10\, {\rm Myr}/1.5\, {\rm kpc}$ at 1 GeV.

\begin{figure}
    \includegraphics[width=\columnwidth]{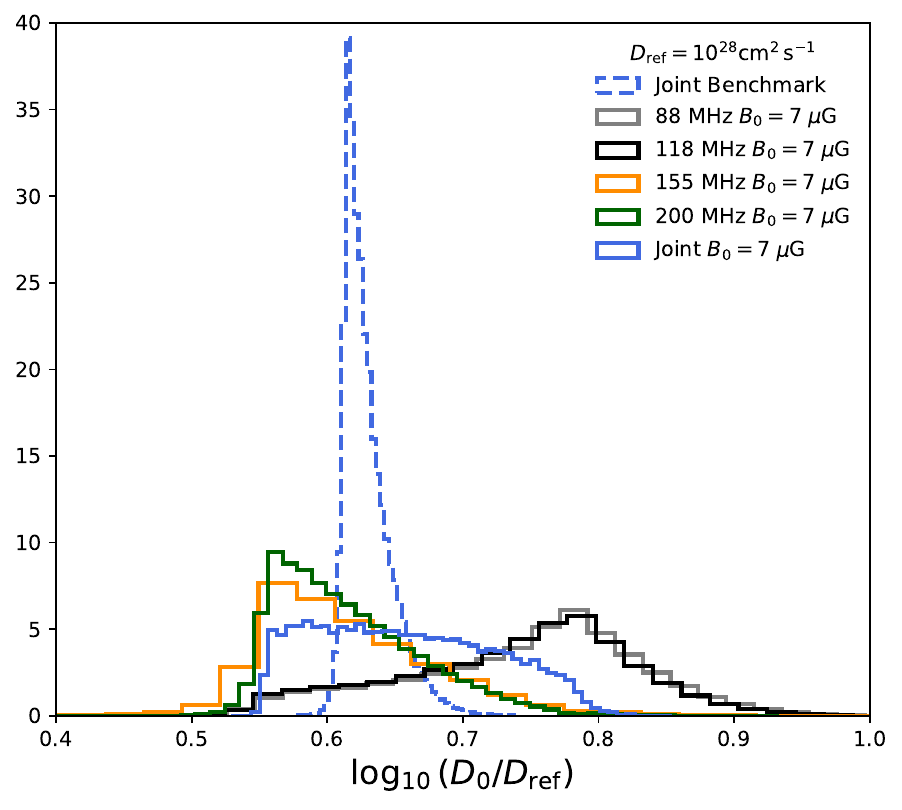}
    \caption{
  Same as in Fig.~\ref{fig:Do_test_zh_cases} but comparing a model with $B_0 = 7 \; \mu G$
    with the benchmark scenario.}
    \label{fig:Do_test_bfield}
\end{figure}

\begin{figure}
    \includegraphics[width=\columnwidth]{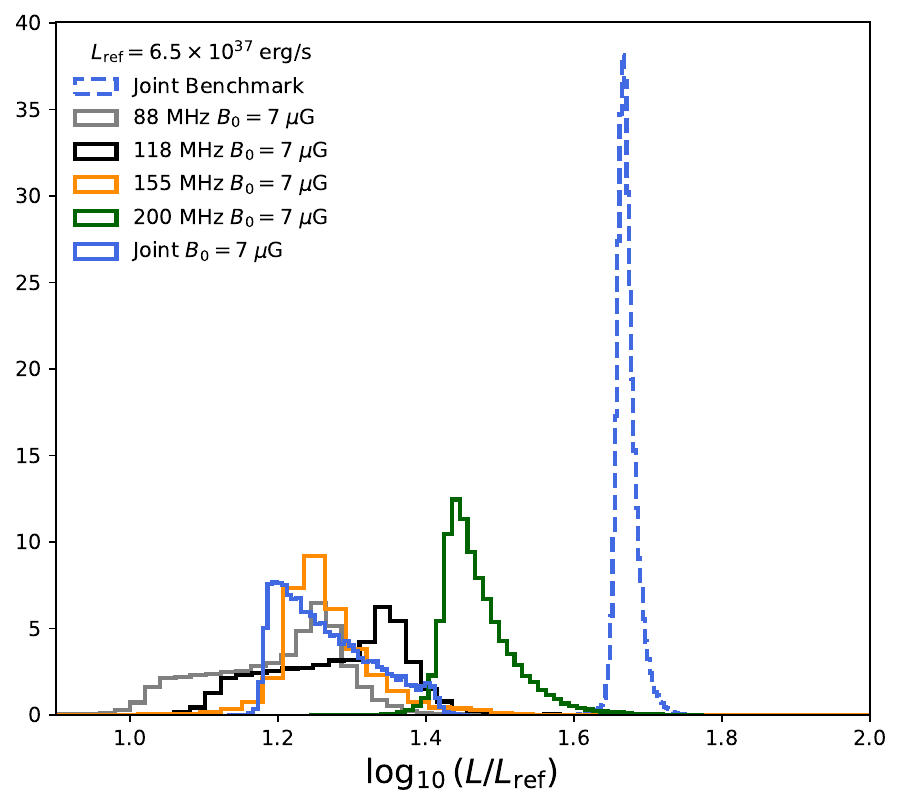}
    \caption{Same as in Fig.~\ref{fig:Do_test_bfield} but for $L$.
    }
    \label{fig:Lo_test_bfield}
\end{figure}

We now consider a different normalization $B_0$ of the magnetic field. For the benchmark model, we have taken $B_0=4.3\,\mu G$, based on the determination obtained in Ref.~\cite{Gaensler:2005} from Faraday rotation measurements.
Other studies have found both smaller values~\cite{2025MNRAS.541.1106S,2024MNRAS.535.1944L} and significantly larger ones~\cite{Seta:2022uoy,2022MNRAS.510...11H}.
To test the robustness of our results, we have considered a value of $B_0=7\,\mu G$, which corresponds to the upper limit on the total magnetic field reported in Ref.~\cite{Mao:2012}.
For this choice and, therefore, also for smaller values, the synchrotron emission is not the dominant source of energy losses for CR electrons in the LMC. Therefore, one expects that rescaling the magnetic-field strength mainly affects the normalization of the synchrotron map rather than its morphology, an effect that can be absorbed into a shift of the luminosity parameter $L$.
Indeed, as shown in Fig.~\ref{fig:Do_test_bfield}, the posterior distributions of $D_0$ for the two choices of $B_0$ are in good agreement. Moreover, the distribution of $L$ shifts to smaller values when $B_0$ is increased, as shown in Fig.~\ref{fig:Lo_test_bfield}, by an amount consistent with the expected $B_0^{2}$ scaling of the synchrotron emissivity.

As a final check, we have changed the CR electron injection index $\alpha_{\rm inj}$ in Eq.~\ref{eq:injection_spectrum}.
The reference choice in Table~\ref{table:params} is $\alpha_{\rm inj}=2.65,$ 
consistent with indexes inferred from CR electron measurements at Earth and from gamma-ray data, typically in the range between $2.6$ and $2.75$~\cite{2011APh....34..528D,Fornieri:2019ddi,DeLaTorreLuque:2025zsv}. 

\begin{figure}[h]
    \includegraphics[width=\columnwidth]{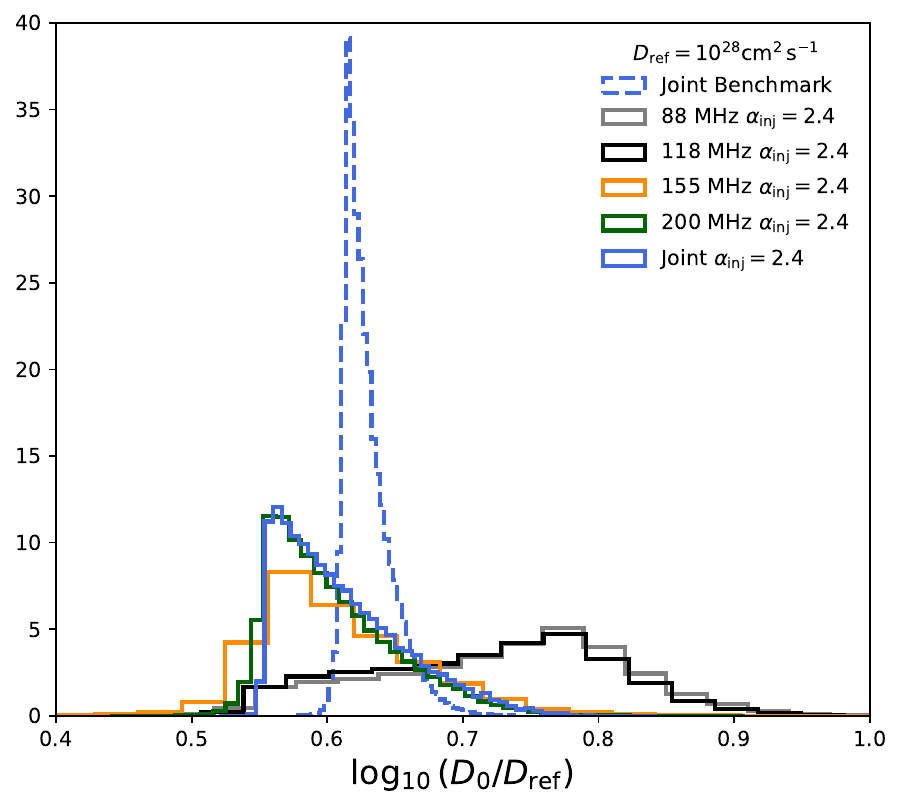}
    \caption{Same as in Fig.~\ref{fig:Do_test_zh_cases} but comparing a model with $\alpha_{\rm{inj}}=2.4$ with the benchmark scenario.
    }
    \label{fig:Do_test_inj_spectrum}
\end{figure}

\begin{figure}[h]
    \includegraphics[width=\columnwidth]{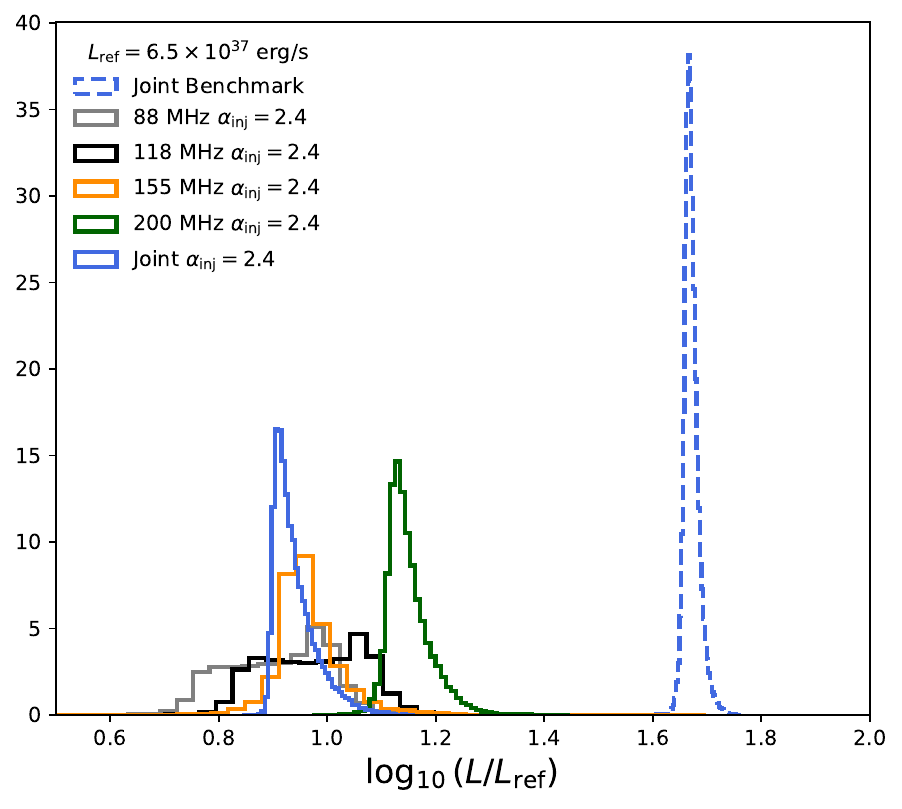}
    \caption{Same as in Fig.~\ref{fig:Do_test_inj_spectrum} but for $L$. 
    }
    \label{fig:Lo_test_inj_spectrum}
\end{figure}

We adopted the same value for the LMC, assuming that it hosts similar CR accelerators as the Milky Way.
However, 
we have also tested the case of a harder energy distribution with $\alpha_{\rm inj}=2.4,$ used for instance in~\cite{Gaggero:2013rya} in the context of a 3D model of the Milky Way.
The impact of adopting a softer spectrum can be understood in an analogous way.
Our results are shown in Figs.~\ref{fig:Do_test_inj_spectrum} and~\ref{fig:Lo_test_inj_spectrum}.
The synchrotron emission at the frequencies considered in our analysis traces a population of CR electrons with energies above approximately 1 GeV. Compared to our reference model, an injection index of $\alpha_{\rm inj}=2.4$ increases the number of CR electrons injected at those energies (see Eq.~\ref{eq:injection_spectrum}), and consequently enhances the synchrotron emission. 
This effect is compensated by a smaller value of the luminosity parameter $L$, as shown in Fig.~\ref{fig:Lo_test_inj_spectrum}. Instead, the distributions of $D_0$ for the two choices of $\alpha_{\rm inj}$ remain in good agreement; see Fig.~\ref{fig:Do_test_inj_spectrum}.

\end{document}